\documentclass[journal=ancac3, manuscript=article, layout=two  column]{achemso}%

\pdfoutput=1
\usepackage{amsmath}
\usepackage{pgfplotstable} 
\pgfplotsset{compat=1.6}

\usepackage{xcolor} %
\usepackage{booktabs} 
\usepackage{tabularx} %
\usepackage{siunitx} 
\usepackage{layouts} %
\usepackage{dblfloatfix}

\graphicspath{{Figures/}} %
\usepackage{subcaption}
\usepackage{scalerel}

\usepackage[version=3]{mhchem} 

\newcommand{\etal}{\textit{et al}.}

\SectionsOn
\AbstractOn

\author{Kusse S. Bersha}
\affiliation[Science Institute]
{Science Institute of the University of Iceland, 107 Reykjav\'{\i}k, Iceland}

\author{Alejandro Pe\~{n}a-Torres}
\affiliation[Science Institute]
{Science Institute of the University of Iceland, 107 Reykjav\'{\i}k, Iceland}

\author{Hannes J\'onsson}
\affiliation[VoN]
{Faculty of Physical Sciences, University of Iceland, 107 Reykjav\'{\i}k, Iceland}
\alsoaffiliation[Aalto University]
{Department of Applied Physics, Aalto University, FI-00076 Espoo, Finland}
\email{hj@hi.is}

\title[From AC-STEM Image to 3D Structure: A Systematic Analysis of Au$_{55}$ nanocluster]
  {From AC-STEM Image to 3D Structure: A Systematic Analysis of Au$_{55}$ nanocluster}

\keywords{Nanocluster, gold, optimization, atomic structure, electron microscopy}

\begin{document}

\begin{tocentry}

\includegraphics[width=8.75cm,height=3.5cm]{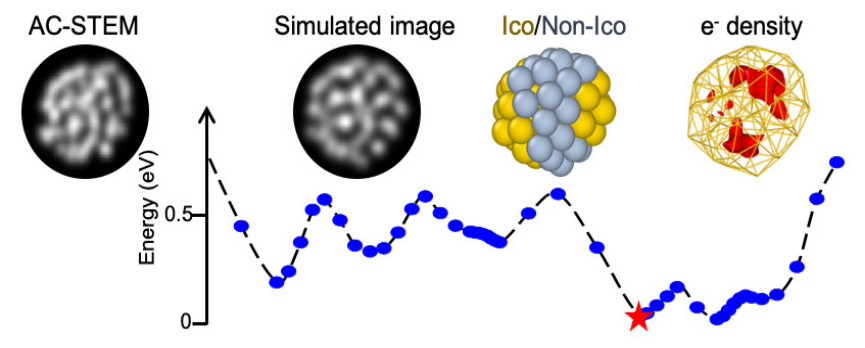}




\end{tocentry}

\begin{abstract}
Aberration-corrected scanning electron microscopy (AC-STEM) can provide valuable information on the atomic structure of nanoclusters, an essential input for gaining an understanding of their physical and chemical properties. A systematic method is presented here for the extraction of atom coordinates from an AC-STEM image in a way that is general enough to be applicable to irregular structures. The two-dimensional information from the image is complemented with an approximate description of the atomic interactions so as to construct a three-dimensional structure and, at a final stage, the structure is refined using electron density functional theory (DFT) calculations. The method is applied to an AC-STEM image of Au$_{55}$. Analysis of the local structure shows that the cluster is a combination of a part with icosahedral structure elements and a part with local atomic arrangement characteristic of crystal packing, including a segment of a flat surface facet. The energy landscape of the cluster is explored in calculations of minimum energy paths between the optimal fit structure and other candidates generated in the analysis. This reveals low energy barriers for conformational changes, showing that such transitions can occur on laboratory timescale even at room temperature and lead to large changes in the AC-STEM image. The paths furthermore reveal additional cluster configurations, some with lower DFT energy and providing nearly as good fit to the experimental image.
\end{abstract}


\section{}

In the past few decades, the study of small metallic nano-clusters has been an active field of research, in part as a result of the numerous applications in optoelectronics, magnetism and catalysis.\cite{Ray_rev_2010,Liu_mag_book_2014,Corain_Cat_Book_2011}  
Knowledge of the atomic scale structure is a prerequisite for gaining a 
clear understanding of cluster properties, as is the case for any form of matter, and can help to identify possible future applications.
Experimental studies of the atomic scale structure of nanoclusters have, however, been hampered by lack of appropriate tools.  
The aberration-corrected scanning transmission electron microscope (AC-STEM) technique\cite{Dellby01,Batson2002} 
has recently become a powerful tool for gaining information about the structure of nanoclusters and has thereby opened up new territory in nanocluster research.\cite{Li2008c,VanAert_2011,Deepak2018} The information extracted from AC-STEM images has, however, so far mainly been
qualitative and analysis techniques are needed for systematic, quantitative structural determination.
       
A great deal of AC-STEM measurements have been dedicated to the study of gold nanoclusters. 
The analysis of the experimental images has mainly involved visual comparison
with pre-calculated images for the three regular structure motifs, \textit{i.e.} decahedral, icosahedral and cuboctahedral geometries.\cite{Li2008c,Wells_Palmer_2015,Jian_Palmer_Au55_2015,Liu_Palmer_2017,doi:10.1002/smll.201002168,Wang_small_2011} 
While the basic structure of many clusters has been identified using this technique, there are also clusters, such as Au$_{55}$, 
where the structure does not fit any of the standard motifs. This is likely also going to be the case for many multicomponent clusters.
Wang and Palmer\cite{Wang2012b} carried out an extensive study of Au$_{55}$, 
and could relate about half the images obtained to a prediction made earlier by Garz\'on \etal\cite{PhysRevB.54.11796,PhysRevLett.81.1600}
based on simulations using an empirical potential function. 
    
While the generation of an AC-STEM image from an assumed structure is typically carried out using electron scattering calculations,\cite{QSTEM}
Li \etal\cite{Li2008c} have shown that the intensity in experimental images of Au$_n$ clusters with n$\leq$1500 increases nearly linearly with the number of atoms. Furthermore, they showed that images generated from electron scattering calculations can be reproduced to good approximation
with a simple superposition of a Gaussian-like contribution from each atom in the cluster. 
This simplifies greatly the simulation of an AC-STEM image from an assumed atomic structure. 
For a given model of the structure, an approximate simulation of an AC-STEM image for a Au cluster is, therefore, quite simple. 
%
%

Finding the stable, low energy structures of a nanocluster theoretically is, however, a significant challenge.
The number of local minima on the energy surface is typically large
and an exhaustive exploration is not practical even for relatively small clusters.\cite{Wales_1999,wales_book2004,Oakley_2013} 
While it may be easy to find some low lying minima, the search for the global minimum may be impossible and 
it is difficult to know whether it has been found or remains to be seen.
This even applies to calculations using simple, empirical potential functions, which have limited predictive power 
(as demonstrated below, but for which the pioneering work of Garz\'on \etal on Au$_{55}$ is a notable exemption), 
while the more accurate, but still typically approximate electronic structure calculations, 
are limited to even smaller clusters and less exploration. 
    
    
Several studies have made use of genetic algorithms (GAs) or other optimization algorithms combined with empirical potential functions 
or electronic structure calculations in order to obtain structures of nanoclusters.
\cite{Watanabe_2001,Kirkland_book_2010,Ferrer_2008,Deepak_2010,Logsdail_2012,Meredig_2013} 
A method tailored to the analysis of AC-STEM data in particular has been presented by Yu \etal\cite{YuMin_2016} who applied it to 
a model of a large Au nanoparticle ($\sim$6000 atoms). 
For such large clusters, the local ordering of the atoms is the same as in the crystal and the analysis can be simplified in terms of positions and intensities of columns and rows of atoms, as well as surface facets. 
However, the structure of smaller Au clusters can be less regular, 
increasing the difficulty of both the search of low energy minima and the analysis of the experimental image.
In fact, it has been shown recently that the common notion of magic clusters, 
{\it i.e.} cluster that are particularly stable because they correspond to 
shell closings of the ideal structure motifs, does not hold for Au clusters in the range of 100 to 2000 atoms.\cite{Garden_2018}



    
        
In this work, a method is presented for the analysis of AC-STEM images that is applicable to disordered as well as ordered clusters. 
An AC-STEM image of a Au$_{55}$ cluster reported by Wang and Palmer\cite{Wang2012b} is analyzed using a two phase optimization method based on an objective function that includes the discrepancy between a simulated and the experimental image as well as an approximate estimate of the energy of the cluster.
In this way, several structures of Au$_{55}$ are generated and finally 
optimized using DFT calculations. 
Additionally, the minimum energy path between the various structural candidates is calculated using DFT to explore further the potential energy surface of the system, thereby revealing new local minima on the energy surface and identifying facile transitions between the structures. 
A brief account of a preliminary version of the analysis method has been published previously.\cite{Sukuta2017}
    


\section{A. Generation of trial structures}
\label{Results}


Using the results obtained by Li \etal \cite{Li2008c} 
showing that an AC-STEM image of a Au cluster can be approximated by a superposition of Gaussians, 
the intensity at a given pixel can be described as a linear combination of intensity contributions from each of the atoms in the cluster. 
Thus the intensity at pixel $(i,j)$ can be written as 
\small
    \begin{equation}\label{eq:image-model}
        I_{ij}(\boldsymbol{x},\!\boldsymbol{y},\!\sigma)\!=\!\sum_{k=1}^{N}\!\mathcal{A}\exp\!{\left(\!-\frac{\!\Big(x'_{i}\!-\!x_{k}\Big)^2\!+\!\Big(y'_j\!-\!y_{k}\Big)^2\!} {2\sigma^2}\right)},
    \end{equation}
\normalsize
where $N$ is the number of atoms in the cluster with coordinates $\boldsymbol{x}\!=\!(x_1,\cdots,x_N)$ and $\boldsymbol{y}\!=\!(y_1,\cdots,y_N)$
projected on a plane parallel to the image, $x'$ and $y'$ are the spatial coordinates corresponding to the pixels of the simulated image, 
and $\sigma$ and $\mathcal{A}$ are the width and height of the Gaussian peaks. 
The parameter $\mathcal{A}$ is obtained by matching the integrated intensity of $N$ atoms to the integrated intensity of the experimental image
%
    \begin{equation}\label{eq:gaussian-amplitude}
        \mathcal{A} = \frac{1}{2\pi \sigma^2 R^2 N} \sum_{i=1}^m \sum_{j=1}^n I'_{ij},
    \end{equation}
where ${I'}$ is the experimental image, $m$ and $n$ give the number of rows and columns of pixels in the image and $R$ is the resolution of the image in pixels/\AA{}. The unknown parameters of the model are the atom coordinates and the width, $\sigma$, of the Gaussians.
These are estimated by fitting the simulated image to the AC-STEM image. 

A simple measure of the agreement between a simulated image and an experimental image can be obtained 
as a sum over the squared difference between the measured, ${I'}$, and calculated, ${I}$, intensity at each pixel
    \begin{equation}
    \label{eq:sum-of-squared-errors}
        \chi^{2}(\boldsymbol{x}, \boldsymbol{y}, \sigma) = \sum_{i=1}^{m}\sum_{j=1}^{n}\Big(I'_{ij} - I_{ij}(\boldsymbol{x}, \boldsymbol{y}, \sigma) \Big)^2.
    \end{equation}
The displacement of the atoms and change in the Gaussian width in the model that reduces the discrepancy 
between the two images most rapidly can be found by analytically 
differentiating $\chi^{2}$, see the $Methods$ section, to make the optimization more efficient.
This simple $\chi^2$ measure turns out, however, to be less than optimal, as shown below. 

At this level of approximation, the AC-STEM image does not give any information about the $z$-coordinate of the atoms. 
Therefore, an objective function that also includes the energy of the cluster is included to complement the fit to the image. 
An objective function is thus defined as the weighted sum of the pixel-by-pixel discrepancy and an estimate of the energy of the cluster as
    \begin{equation}\label{eq:objective-function}
        f(\boldsymbol{x},\boldsymbol{y}, \boldsymbol{z}, \sigma) = \omega\chi^2(\boldsymbol{x}, \boldsymbol{y}, \sigma) + U(\boldsymbol{x}, \boldsymbol{y}, \boldsymbol{z}),
    \end{equation}
where $\boldsymbol{z}\!=\!(z_1,\cdots,z_N)$ corresponds to the $z$-coordinates of the atoms, $U$ is the estimated energy of the cluster
and $\omega$ is a weight parameter used to control the relative importance of the two contributions. 
The gradient of this combined objective function then consist of the gradient of the energy (the negative of the force acting on the atoms) 
plus a scaled contribution from the gradient of $\chi^2$. 
Setting $\omega\!=\!0$ during optimization results in pure energy minimization and identifies a low energy structure but not a fit to the AC-STEM image. On the other hand, setting $\omega$ to a large value results in pure image fitting providing two-dimensional images that fit the experimental image well, but likely correspond to a structure with high energy. 
Therefore, by choosing an intermediate value for the weight, a compromise between image fitting and energy is struck. 
The energy is estimated here using the effective medium theory (EMT) potential function,\cite{JACOBSEN1996394} 
which works well in the present context even though the lowest energy structure found for Au$_{55}$ is predicted to be entirely different from
the estimate obtained from the analysis of the AC-STEM image, as shown below. 
It is important that the method does not require a highly accurate estimate of the energy,
as it mainly serves to control the nearest neighbor distance between the atoms and to provide interatomic attraction to
ensure that a compact cluster model is generated.
 
\begin{figure}[!t]
        \centering
        \includegraphics[width=\columnwidth]{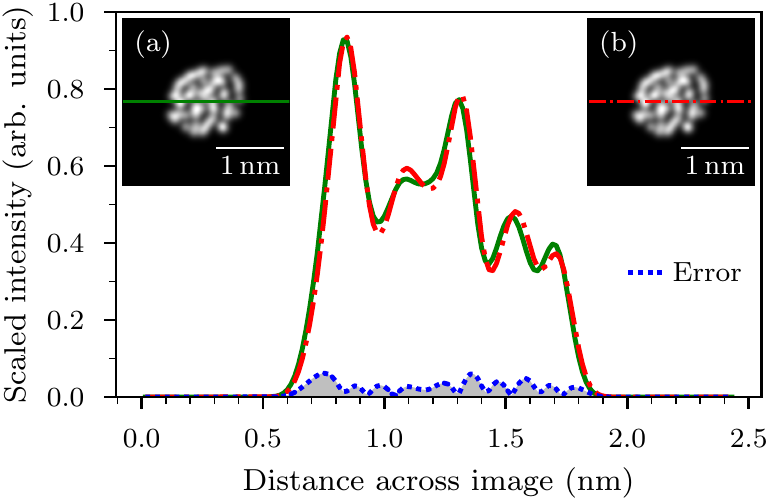}
        \caption{
        (a) Experimental AC-STEM from ref.\cite{Wang2012b} and  
        (b) simulated image of a Au$_{55}$ cluster after the first phase 
        of the fitting procedure, as well as line scans of the intensity (green for experimental and broken red for simulated). 
        At this point the energy of the cluster is high as the $z$-coordinate of the atoms has not been optimized, except for local minimization.
        The difference between the two intensity profiles is shown by the blue dotted curve.}
        \label{fig:intensity-profile} 
    \end{figure}

The method for generating the atomic coordinates of the nanocluster consists of two phases.
The first phase starts out by locating local intensity maxima in the AC-STEM image. 
The first atom in the model is then placed at the location of the highest maximum and its $z$-coordinate is arbitrarily given a value of $z\!=\!0$. 
The intensity contribution of this atom is then subtracted from the experimental image and a second atom is placed at the location of the maximum in the reduced image. 
The $z$-coordinate of the second atom, and all subsequent atoms, is generated randomly between $-\SI{10}{\angstrom}$ and $\SI{10}{\angstrom}$ while ensuring a minimal interatomic distance of $\SI{2.8}{\angstrom}$. 
Local minimization of the combined objective function is performed 
with a weight of $\omega\!=\!1000$ eV after the 
placement of each atom. This process is continued until the coordinates of a predefined number, $N$, of atoms has been assigned. 
By repeating this process with different random number seeds, several three-dimensional atomic structures are generated with ($x,y$) coordinates that almost perfectly fit the AC-STEM image, but have relatively high energy and 
are slightly elongated in the $z$-direction.
They serve as input for a structure optimization procedure in the second phase. 
The optimal width of the Gaussians, $\sigma$, is determined in this first phase of the fitting and is 
kept fixed during the optimization of the structures in the second phase. 
Figure \ref{fig:intensity-profile} shows a comparison between the experimental image (inset $a$)  and an image obtained by the fitting process described above (inset $b$). 
Horizontal line scans illustrate the excellent agreement with respect to both the peak positions and the relative peak intensities. 
%
    {}    

\section{B. Optimization of the structures}

\begin{figure}[!t]
        \centering
        \includegraphics[width=\columnwidth]{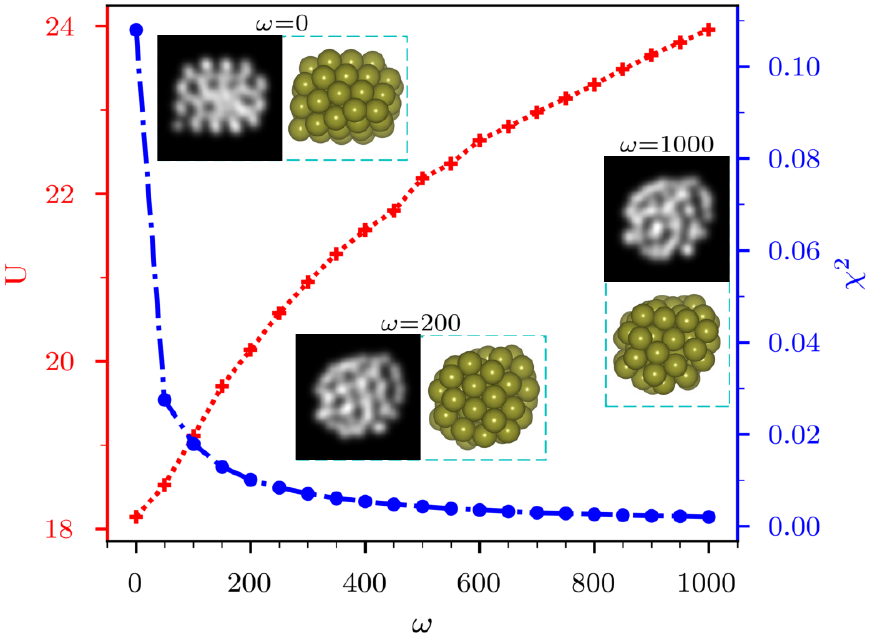}
        \caption{
        Relationship between the energy, $U$, of a Au$_{55}$ cluster and the measure, $\chi^2$, of the fit to an AC-STEM image 
        as a function of the weight parameter $\omega$ (in eV) in eq~\ref{eq:objective-function}. 
        The axis on the left and the red curve show the energy given by the EMT potential function, 
        the axis on the right and the blue curve show the sum of squared discrepancy, $\chi^2$, between the experimental AC-STEM image 
        (shown in figure 1(a)) and a simulated image. 
        Note the large discrepancy in the image generated from the lowest energy configuration obtained from the EMT potential, 
        where the local ordering is locally that of a face centered cubic crystal, 
        but this approximate interatomic potential still suffices for the present purposes.
        }
        \label{fig:emt-rss-omega}
    \end{figure}

An optimization of the atomic coordinates of the cluster generated in the first phase is carried out by minimizing the combined objective function given by eq \ref{eq:objective-function} with a smaller value of the weight, $\omega$, on the image. 
In order to get a good compromise between the information from the AC-STEM image and the energy estimate, 
a suitable value of $\omega$ needs to be determined. Figure \ref{fig:emt-rss-omega} shows the relationship between image fit and cluster energy for different values of the weight parameter. When $\omega$ is small, the objective function is dominated by the energy term and the optimization therefore leads to low energy structures, but there is not much similarity between the experimental image and the fit, as shown in Figure \ref{fig:emt-rss-omega} for $\omega\!=\!0$ eV. Conversely, when $\omega$ is large, the simulated image fits well the experimental image but the resulting cluster structure has relatively high energy and a rough surface as shown in Figure \ref{fig:emt-rss-omega} for $\omega\!=\!1000$ eV.
%
        {} 
%
    
We have used two different methods in the structure optimization in the second phase.
Given that a value for $\omega$ ranging between $100$ and $300$ eV seems to give a good compromise between the image fit and the energy, 
the optimizations are carried out for different values within this range. The two methods used give roughly the same success rate
with similar computational effort and are briefly 
described below, but more details can be found in the $Methods$ section.
    

\subsection{A. Optimization via saddle point searches}    

This method is based on finding a new local minimum on the objective function surface by 
driving the system through regions of first order saddle points and then sliding down on the other side. 
The method is referred to as global optimization using saddle traversals (GOUST).\cite{Plasencia_2014} 
For every cluster structure obtained in the first stage, multiple saddle point searches are performed.
The minimum mode following method is used to converge onto the saddle points.\cite{Henkelman_1999,Gutierrez_2017} 
Afterwards, a displacement of the system along the minimum mode at the saddle point is carried out, 
followed by minimization to slide down to the adjacent minimum. 
In this way, the system heads down a funnel in the combined objective function surface,
by hopping over barriers, until a minimum with the lowest value compared to nearby local minima
has been reached.

\subsection{B. Optimization using a genetic algorithm}

Alternatively, a genetic algorithm (GA) is used, as is commonly done in nanocluster simulations.\cite{DAVEN1996, Michaelian:1999, Vilhelmsen:2014, YuMin_2016, VandenBossche:2019} 
New structures are generated from parent structures by crossover and mutations. To maintain population diversity and guarantee the survival of the fittest, parent structures are selected with a bias towards configurations that give low values of the objective function and have been less frequently selected.
    
Both of these optimization methods provide a set of structures that have low values of the objective function and are thereby candidates for
describing the atomic structure of the cluster. 
After the GOUST or GA optimization, local minimization calculations of the objective function are carried with the weight parameter 
reduced gradually from $\omega\!=\!100$ eV 
to $\omega\!=\!50$ eV.
In this way, the energy of a cluster is gradually reduced while maintaining some level of a fit to the experimental AC-STEM image.
At this point, the EMT part of the objective function is replaced by a DFT calculation using the PBEsol functional to get a more accurate
estimate of the energy and a minimization is carried out with $\omega\!=\!50$. 
Since DFT calculations involve significant computational effort, 
only the best
structures are selected for this final minimization.
Finally, the contribution of the AC-STEM image to the objective function is turned off, $\omega\!=\!0$ eV, 
and the DFT energy minimized.


\section{Results and Discussion}

     \begin{figure}[!b]
        \centering
        \includegraphics[width=\columnwidth]{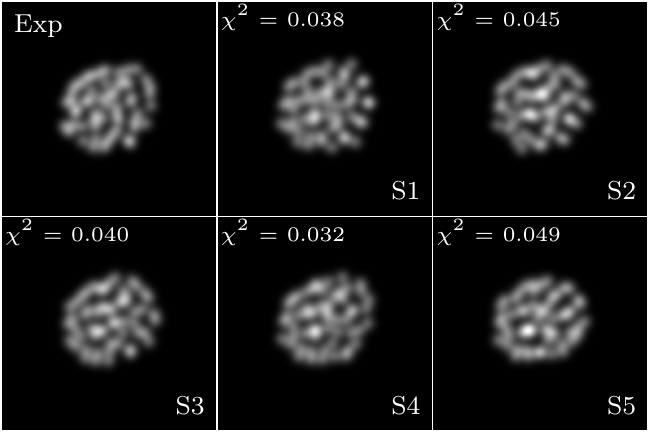}
        \caption{
        The AC-STEM image obtained from ref.~\cite{Wang2012b} and simulated images, labeled $S1$-$S5$, 
        for five of the Au$_{55}$ cluster structures obtained from the two phase fitting process 
        which concludes with energy minimization using DFT/PBEsol. 
        The $\chi^2$ value is the sum of squared pixel intensity differences between the  
            AC-STEM image and the simulated image for each structure.
            }
        \label{fig:simulated-images}
    \end{figure}
  
Figure \ref{fig:simulated-images} shows a set of simulated images (labeled $S1-S5$) obtained by fitting an AC-STEM image
reported by Wang and Palmer.\cite{Wang2012b} 
The $\chi^2$ value shows the sum of squared pixel intensity differences between the simulated image and the experimental one,
as defined by eq \ref{eq:sum-of-squared-errors}.
The main features of the experimental image are reproduced quite well, in particular the bright feature in the lower left hand region and 
the ring surrounding it. Structure $S4$ has the lowest $\chi^2$ value. 
Nevertheless, some shortcomings are also evident, mostly on the right hand side of the image. 
The $S1$ structure, which has only slightly higher $\chi^2$ value,
shows some inconsistencies at the edges of the image.
The simulated image has distinct peaks in the intensity along the edge while the experimental image is more diffuse in this region possibly
because of large thermal vibrations of the surface atoms. 
The reason for such discrepancies can also be transitions between structures during the imaging.
In the experiments, the temperature is raised beyond room temperature because of heating by the electron beam, and
observations of transitions during the imaging have been reported. 
\cite{Wang_small_2011,Wang2012b,Wang_prl_2012}
This can make it difficult to obtain close correspondence between any one cluster structure and an experimental image
especially for such a small and disordered cluster.
Considering such limitations, the agreement between these simulated images and the AC-STEM image can be considered to be
adequate.

    
\subsection{Minimum energy paths}

\begin{figure*}[!t]
    \centering
    \includegraphics[width=0.9\textwidth]{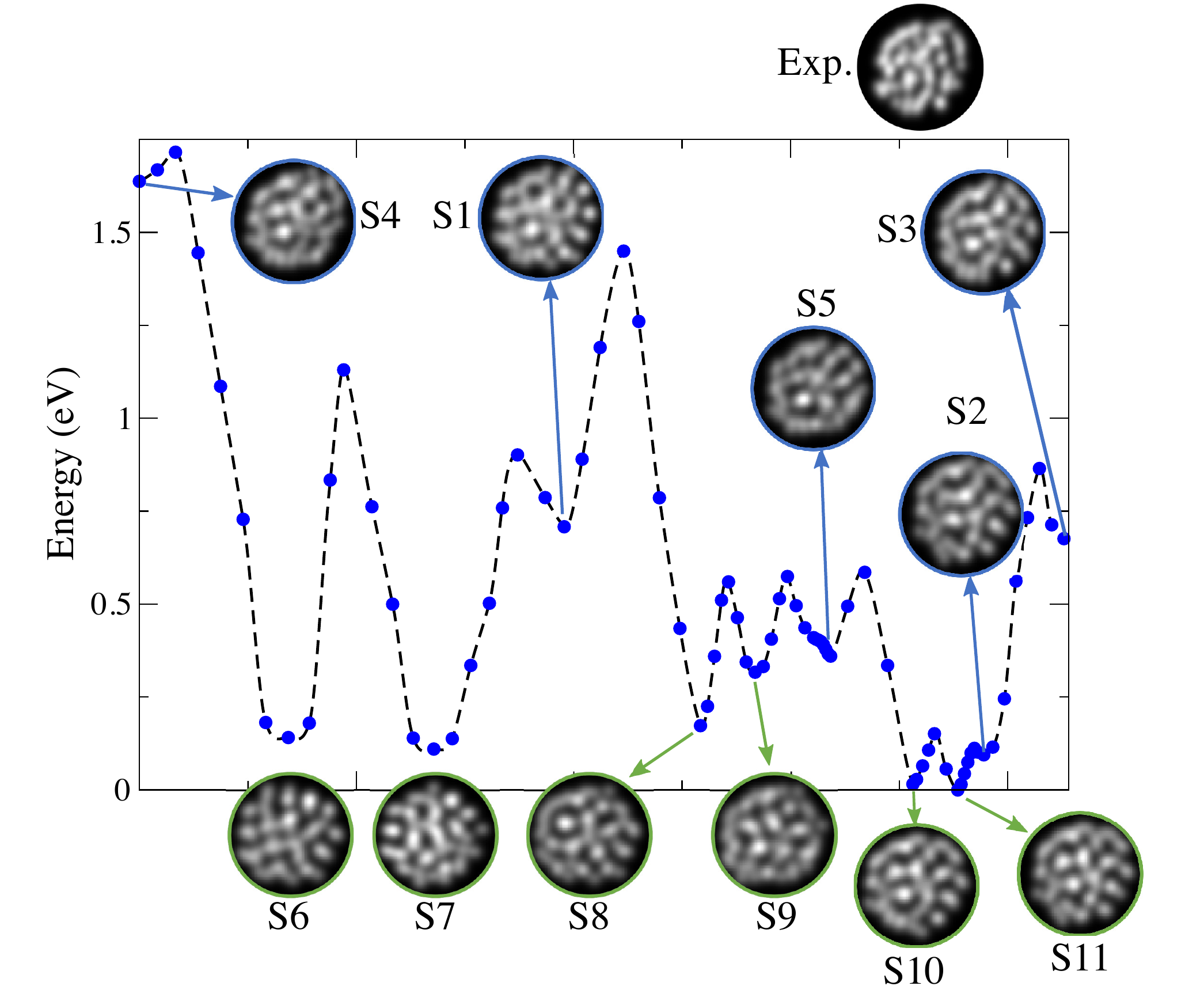}
    \caption{
    Minimum energy path between the five optimal Au$_{55}$ structures, $S1-S5$, obtained in repeated fits of the AC-STEM image. 
    Each blue dot corresponds to a cluster structure (an image) along the path obtained from the nudged elastic band calculation. 
    New local minima, $S6$-$S11$, are found along the path and the corresponding simulated images are shown below. 
    For comparison, the  AC-STEM image is located above the graph.
    }
    \label{NEB_barriers}
 \end{figure*}

To explore the energy landscape in the vicinity of the optimal structures, $S1-S5$, 
the minimum energy path connecting the corresponding minima on the DFT energy surface is calculated.
The cluster structures are first ordered according to the average total distance between the atoms 
(\textit{i.e.} the root mean square displacement, RMSD) and the path between adjacent pairs then 
calculated using the climbing image nudge elastic band (CI-NEB) method.
\cite{Henkelman_2000a,Henkelman_2000b,Asgeirsson_2020} with atomic forces obtained from DFT. 
When points indicating intermediate minima are found along the path, the corresponding configurations are relaxed 
separately to the local minima, 
thereby revealing new low energy configurations, $S6-S11$, as can be seen in figure \ref{NEB_barriers}. 
Interestingly, minimum energy structures that are adjacent on the path can give quite different simulated images, 
showing that significant structural changes can take place as the system hops over a single energy barrier.

%

Table 1 gives the value of the $\chi^2$ measure of the fit to the AC-STEM image for the various structures, 
as well as the cluster structure with lowest energy found from the EMT potential function.
Remarkably, the image produced from the EMT predicted structure ranks rather high even though it bears no resemblance to the AC-STEM
image. There, the local ordering of the atoms corresponds to that of the face centered cubic (FCC) crystal.
It is, therefore, clear that the $\chi^2$ measure of the goodness of fit as defined in 
eq \ref{eq:sum-of-squared-errors} is not sufficiently reliable. 
This sum over all pixels can be dominated by regions of the image that are not revealing about the atomic structure.
%
To tackle this, a modified measure is defined by increasing the weight of the brighter pixels in the AC-STEM image.
The pixels are ranked according to the intensity and divided into two equally large groups, the brighter and the fainter pixels.
A weight parameter is introduced in the pixel-by-pixel discrepancy and equation \ref{eq:sum-of-squared-errors} rewised as
    \begin{equation}\label{eq:weighted-sum-of-squared-errors}
        _{s}\chi^{2}(\boldsymbol{x}, \boldsymbol{y}, \sigma) = \sum_{i=1}^{m}\sum_{j=1}^{n} 
          g_{ij}  \Big(I'_{ij} - I_{ij}(\boldsymbol{x}, \boldsymbol{y}, \sigma) \Big)^2,
    \end{equation}
where $g_{ij}=5$ if pixel $i,j$ is among the brighter pixels, otherwise $g_{ij}=1$.

    
    
Table \ref{tab:chi2} shows how the ranking of the fits changes by placing higher weight on the brighter pixels.
In addition to the structures obtained from the image fit, structures obtained 
from the minimum energy path calculation as well as the Garz\'on structure (referred to as Gar for short) 
and the EMT minimum energy structure
(shown in figure \ref{fig:emt-rss-omega}) are evaluated.
With the increased weight on the brighter pixels, a better ranking of the structures is obtained.
The EMT structure is, for example, now ranked lower than the fitted structures. 
The Garz\'on structure, though, still ranks even lower than the clearly incorrect EMT structure.
 
    \begin{table}[!t]
    \caption{
    Ordering of several Au$_{55}$ structures according to the level of agreement between the simulated and AC-STEM image, 
    using the sum of squared pixel-by-pixel discrepancy measure, $\chi^2$, defined in eq \ref{eq:sum-of-squared-errors}
    and a revised version where the brighter half of the pixels in the AC-STEM image contribute 
    five times more to the sum than the fainter pixels, $_{s}\chi^2$, as defined in eq \ref{eq:weighted-sum-of-squared-errors}.
    By increasing the weight of the brighter pixels, the numerical measure is more consistent with a visual inspection, 
    lowering, for example, the rank of the image generated from the EMT lowest energy structure which clearly does not correspond well to the
    AC-STEM image. 
    }
    \label{tab:chi2}
\resizebox{\columnwidth}{!}
{
\begin{tabular}{|cc|cc|cc}
\hline
$\chi^2$ & Structure  & $_{s}\chi^2$ & Structure \\
\hline
0.032  & S4    & 0.232    & S4    \\
0.038  & S1    & 0.285    & S1   \\
0.040  & S3    & 0.305    & S3   \\
0.043  & EMT   & 0.311    & S5  \\
0.045  & S2    & 0.316    & S10   \\
0.046  & S10   & 0.393    & S2   \\
0.049  & S5    & 0.441    & EMT  \\ 
0.060  & S7    & 0.591    & GAR   \\
0.075  & Gar   & 0.702    & S7   \\
\hline

\end{tabular}
}
    \end{table}

It is clear that the $\chi^2$ measure can be deceiving 
and that by increasing the importance of the brighter pixels in the experimental image, as expressed in eq~\ref{eq:weighted-sum-of-squared-errors}, 
the ranking becomes more consistent with visual inspection (see $_{s}\chi^2$ values in table \ref{tab:chi2}). 
%
It is important to find a reliable numerical measure for the fit 
and while this increase in the weight of bright pixels clearly is a step in the right direction, more
work xon this is needed, possibly by making use of techniques developed in signal processing and related fields.
One advantage of the $\chi^2$ and $_{s}\chi^2$ measures is that they are differentiable so the atom displacements that decrease the
discrepancy measure can, therefore, be obtained from the gradient with respect to the atom coordinates.
    
    \begin{figure}[!b]
        \centering
        \includegraphics[width=0.8\columnwidth]{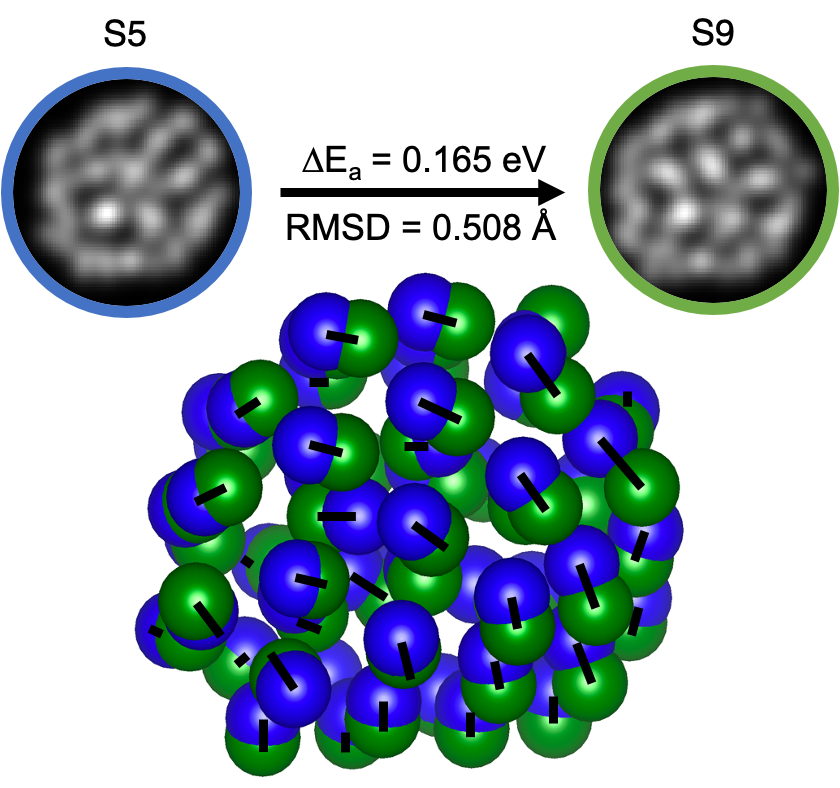}
        \caption{
        Transition between $S5$ and $S9$ structures of the Au$_{55}$ cluster. 
        Top: Simulated images as well as the activation energy and the root-mean-square displacement between the structures. 
        Bottom: A superposition of the atomic structures of $S5$ (blue) and $S9$ (green). 
        Black lines connect initial and final positions of each atom.
        }
        \label{fig:concertedDisp}
    \end{figure}

As can be seen from figure \ref{NEB_barriers}, a large difference can be seen between images of structures that are adjacent on the 
minimum energy path, 
{\it i.e.} separated by a single energy barrier. This occurs because of a concerted displacement of a large number of atoms. 
Note that overall rotation of the cluster is suppressed in the calculations.\cite{Melander_2015}  
To illustrate this, figure \ref{fig:concertedDisp} shows the atomic displacement that take place when $S5$ transforms to $S9$. 
The RMSD between these structures is around 0.5 \AA{}, and the energy barrier that needs to be surmounted is only 0.17 eV, 
which means that these kinds of fluctuations can occur frequently during on time scale of the AC-STEM measurement. 
Figure \ref{fig:concertedDisp} shows a superposition of the $S5$ structure overlapped with the $S9$ structure 
and the displacement of each atom between the two is marked. 
Further analysis shows that the atoms moving most during the transition are located almost exclusively in the region 
corresponding to the right hand side of the image, 
the region is where the simulated images differ most from each other and from the AC-STEM image, as noted above. 

In the experimental study of Wang and Palmer, a time series is reported where the AC-STEM image changes strongly with 
time and they ascribed this to rotations of the cluster.\cite{Wang2012b} 
There, each experimental image was matched as closely as possible to 
the structure predicted by Garz{\'o}n {\it et al.} by rotating the cluster in small steps and 
visually comparing the simulated image to the experimental one. 
The results presented here from the minimum energy path calculation show that the structural fluctuations observed in the experiments 
may in fact be due to concerted change in the relative position of the atoms as the system undergoes a thermally activated structural transition, 
even a hop over just a single energy barrier, rather than rotation of the cluster.
    

    \begin{table}[!b]
    \caption{
    Energy (eV) obtained with DFT/PBEsol of the Au$_{55}$ structures generated
    relative to the most stable one, $S11$. 
    The energy difference between $S10$ and $S11$ is insignificant.
    } 
    \label{tab:energy}
\resizebox{\columnwidth}{!}
{    
     \begin{tabular}{|cc|cc|cc|}
\hline
Structure & $\Delta E$  & Structure & $\Delta E$  & Structure & $\Delta E$  \\
\hline
S11  &  0.00  & S9    &  0.32  & EMT  & 1.12 \\
S10  &  0.02  & S5    &  0.41  & S4    & 1.64 \\
S2    &  0.10  & Gar & 0.51  & Ico  & 1.71 \\
S7    &  0.12  & S3    & 0.68   & Dec & 2.59 \\
S6    &  0.16  & S1    & 0.71  & Cub & 3.31 \\
S8    &  0.18  & DFTB & 0.87 & & \\
\hline

 \end{tabular}
}
\end{table}


Table \ref{tab:energy} gives the relative energy of various cluster structures obtained in the calculations. 
For the sake of comparison, the table also gives results for the high symmetry Au$_{55}$ clusters such as the icosahedral, decahedral and cuboctahedral structures as well as configurations originating from energy minimization using Gupta function, EMT function and DFT based tight binding (DFTB).\cite{VdBossche_2019} 
All these structures have been optimized by minimizing the energy at the DFT level of theory using a small step size to ensure convergence to the nearest local minimum. 
Only minor changes in the structures occur during the DFT calculation for the Gupta, EMT, and DFTB structures, and virtually 
no structural changes occur for the high symmetry structures.
(Information on the changes in the atomic coordinates during the DFT minimization is given in the Supporting Information, figure S1.)

As one might expect, there is a large difference in the energy of the various clusters. The optimal structure found by Garz\'on \etal \cite{PhysRevLett.81.1600} using the Gupta potential has the ninth highest energy of $\SI{0.51}{\electronvolt}$ above the lowest energy structure. The slightly distorted FCC-type structure predicted by the EMT potential function 
is higher, \SI{1.1}{\electronvolt}. The symmetric icosahedral (Ico), Ino-decahedral (Dec) and cuboctahedral (Cub) configurations are the least stable structures. The difference in the energy of the isomers found here from the fit to the AC-STEM image is less than $\SI{0.75}{\electronvolt}$, with the exception of $S4$, and the energy barriers between many of them are small,
as can be seen from figure 4. As mentioned above, the small energy barriers between the isomers may be overcome by thermal activation and lead to the superposition of different isomers during AC-STEM imaging. 
This could account for smearing of the experimental image, especially around edges of the cluster.


\subsection{Structural Analysis}
To characterize the atomic ordering of the clusters, the coordination number of the atoms is calculated and 
the local environment of pairs of atoms classified using the common neighbor analysis (CNA).
Also, the electronic density obtained from the DFT calculations is analyzed. 
More details about the analyses can be found in the $Methods$ section.

\begin{figure}[!b]
    \centering
    \includegraphics[width=\columnwidth]{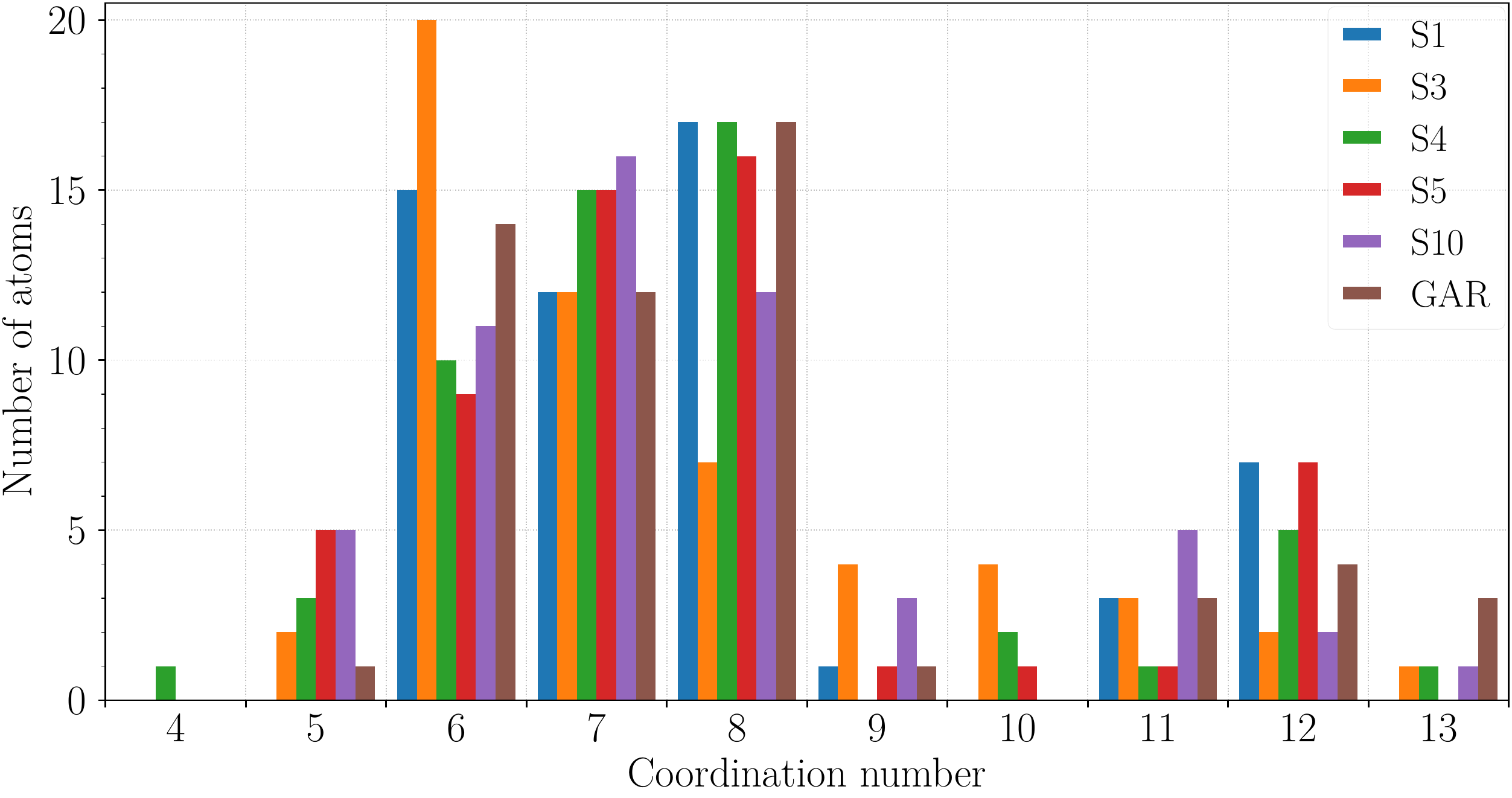}
    \caption{
    Top: Coordination numbers for the five structures that give best fit to the AC-STEM image 
     according to the $_{s}\chi^2$ measure (see table 1), as well as the Garz\'on structure 
     (after local minimization with DFT/PBEsol).
            Vertical axis shows number of atoms found with each coordination number.
            The cutoff distance for defining neighbors (`bonds') is 3.3 \AA{}. 
    }
    \label{fig:crd}
\end{figure}

\subsubsection{Coordination Numbers}
The coordination numbers of the atoms in the five cluster structures 
that best agree with the AC-STEM image according to the $_{s}\chi^2$ measure (see table 1)
are presented in figure \ref{fig:crd}. A cutoff distance of 3.3 \AA{} is used to define neighboring ({\it i.e.} `bonded') atoms. 
A similar analysis for other cluster structures is presented in figure S2 in the Supporting Information.
Most of the clusters show pronounced peaks for coordination numbers 12, 8, 7, and 6. 
The exception is $S10$, the one with lowest DFT energy, which has a peak at a coordination number of 11
(the same is true for the similar $S11$ structure). 
The Garz\'on structure differs from the structures obtained from the AC-STEM fitting in that it has higher coordination numbers, 
in particular three atoms with a coordination number of 13. 
The lower coordination numbers of structures obtained here by fitting the AC-STEM image, 
and even more so the lowest DFT energy structures, 
show that they are less compact than the Garz\'on structure. 


\begin{table*}[!h]
\begin{center}
\caption{
Relative abundance of selected CNA pairs (in \%) for the five structures that give best fit to the AC-STEM image 
according to the $_{s}\chi^2$ measure (see table 1). 
For comparison, the CNA pairs for Garz\'on, icosahedron, EMT, cuboctahedron and Ino-decahedron 
locally minimized with DFT/PBEsol are also given. 
Indices with small relative abundance, such as 444, are ignored. 
}
\label{table:cnatable}
\small
\begin{tabular}{|c|cccccccccc|cccc}
  \small


CNA index & S1  & S3 & S4 & S5 & S10 & Gar   & Ico& EMT & Cub & Dec  \\

\hline

       211    & 3    & 9   & 3   & 4     & 2    & 4        &    0  &  9    & 33   & 18  & \\
       311    & 31  & 28 & 31 & 31   & 27  &  27     &  26  &  37  & 11   &  14 & \\
       322    & 24  & 12 & 21 & 17   &  21 &  24     &  26  &  0    &   0   &   9  & \\
       421    & 2    & 6   & 4   & 1     &  3   & 3        &   0   &  44  & 56   &  23 & \\
       422    & 17  & 6   & 14 & 13   & 11  &  21     &  38   &   0  &    0  &  30 & \\
       433    & 6    & 9   & 4   & 8     &  6   &  4       &    0   &   0  &    0  &    0 & \\
       544    & 2    & 3   & 3   & 4     &  2   &  2       &    0   &   0  &    0  &    0 & \\
       555    & 3    & 1   & 2   & 2     &  3   &  4       &  10   &   0  &    0   &   2 & \\ 
\hline
%
\normalsize
\end{tabular}
\end{center}
\end{table*}


\subsubsection{Common Neighbor Analysis}

The CNA method characterizes the local environment of pairs of atoms by a set of three integer indices. The first one denotes the number of atoms that are neighbors of both atoms in the pair, the second index is the total number of bonds between the common neighbors, and the third index is the number of bonds in the longest continuous chain of bonds between the common neighbors. 
Additional information about CNA can be found in refs.~\cite{Honeycutt1987,Clarke1993, FAKEN1994}

Table \ref{table:cnatable} shows the relative abundance of bonded pairs with the most common CNA indices for 
the five cluster structures that give best agreement with the AC-STEM image according to the $_{s}\chi^2$ measure (see table 1).
Data on additional structures is provided in the Supporting Information.
The CNA signature can, in particular, be used to distinguish between the local ordering in an FCC crystal, 
as in the cuboctahedral structure, and the non-crystallographic, five-fold symmetry icosahedral structure. 
A distinction between the two can be made by looking at the number of pairs of type 
421, 422 and 555.
The main difference between these structures is that pairs of type 421 are
abundant in the cuboctahedral structure but are not present in the icosahedral structure. 
On the contrary, pairs of type 555 and 422 are abundant 
in the icosahedral structure but are not present in the cuboctahedral structure. 
The Ino-decahedral cluster has slightly more 422 pairs than 421 pairs, due to local hexagonal close pack (HCP) ordering at the stacking
faults. Otherwise, it has similar local ordering as the cuboctahedral structure.
The 544 and 433 pairs are indicative of distorted icosahedral order as they form when a single bond is broken in the icosahedron.
The sum of 555, 544, and 433 pairs in some of the structures
obtained from the fit to the AC-STEM image is comparable to the number of 555 pairs in the perfect icosahedron. 
For example, the sum of the 555, 544, and 433 pairs for $S10$ is 10.5\%, 
nearly equal to the total number of 555 pairs in the icosahedral structure, 10.3\%. 
With this information, it becomes clear that the cluster structures obtained here from the AC-STEM image mainly have local atomic 
ordering that is characteristic of the icosahedron. 
Nevertheless, there is also a significant but smaller number of 421 pairs indicative of local FCC-type ordering of the atoms.

%
\begin{figure}[!b]
    \centering
    \includegraphics{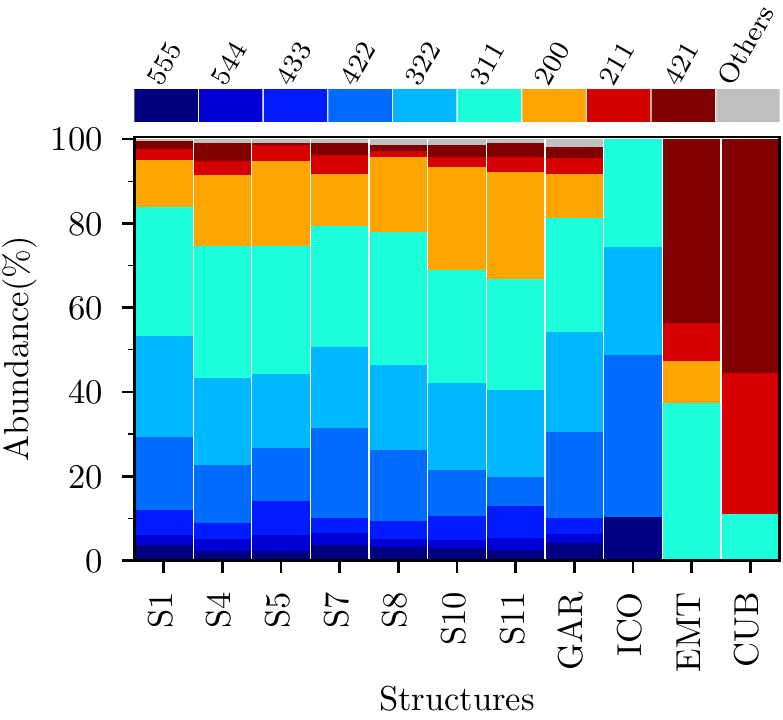}
    \caption{
    Histograms of various CNA pairs for selected structures. The color code is chosen to distinguish between icosahedral local order (blue)  
    and FCC-type local order (red). 
    CNA indices with small relative abundance are included as `others'.
    }
    \label{fig:cna-bar}
\end{figure}

Figure \ref{fig:cna-bar} shows a 
a histogram of the CNA pairs. 
The blue color represents icosahedral while the red represents local FCC ordering as in the cuboctahedral structure.
The main result of the CNA is that the Au$_{55}$ structures found here from the AC-STEM image have local atomic ordering that corresponds 
mainly to distorted icosahedral order, but also some that correspond to local FCC order.
The Garz\'on structure has similar characteristics\cite{PhysRevB.54.11796}  but it has more icosahedral order than the
structures that best fit the AC-STEM image.

\begin{figure}[!t]
    \centering
    \includegraphics[width=\columnwidth]{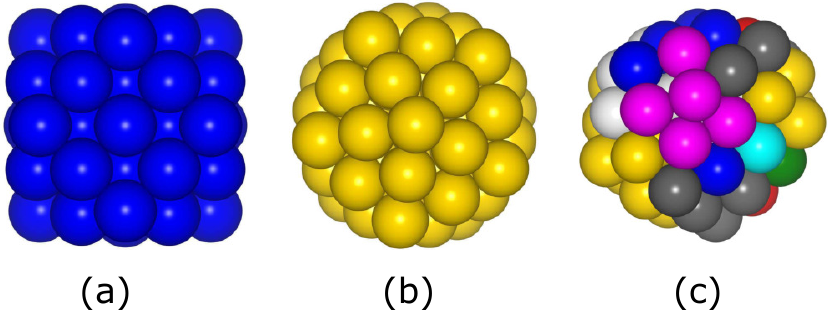}
    \caption{
Au$_{55}$ structures with atoms colored according to three of the CNA pairs: 421, 422 and 555. 
(a) cuboctahedral (Cub), (b) Icosahedral (Ico) and (c) $S10$ structure. 
Blue atoms are involved in 421 pairs, yellow atoms are involved in 422 and 555 pairs, magenta atoms are involved in 421 and 422 pairs
(a characteristic of local hexagonal close pack ordering, HCP). Red atoms are characteristic of local FCC ordering, where 421 pairs are formed but not 422 pairs. White atoms are involved in all three types of pairs, 421, 422 and 555, and gray atoms are not involved in any of those three.}
    \label{fig:CNA_example}
\end{figure}{}

To illustrate this analysis further, figure \ref{fig:CNA_example} shows color-coded configurations of the icosahedral, cuboctahedral and $S10$ 
structures. Even though the $S10$ structure appears to be highly disordered, it presents mostly an icosahedral-like structure 
with just a few atoms containing pairs related to local FCC ordering. 
Every atom in the cuboctahedral structure forms at least one 421 bonded pair but no 422 or 555 pair. 
Every atom in the icosahedral structure forms at least one 422 and one 555 pair but no 421. 
In the $S10$ structure, the majority of atoms have at least one 555 and one 422 pair, but some atoms are involved in 421 pairs.
These three types of pairs, 421, 422 and 555, can be used to classify the local atomic ordering in the core of the cluster, 
but the outer shell atoms are mainly involved in pairs with lower CNA index.

\begin{figure}[!t]
    \centering
    \includegraphics[width=\columnwidth]{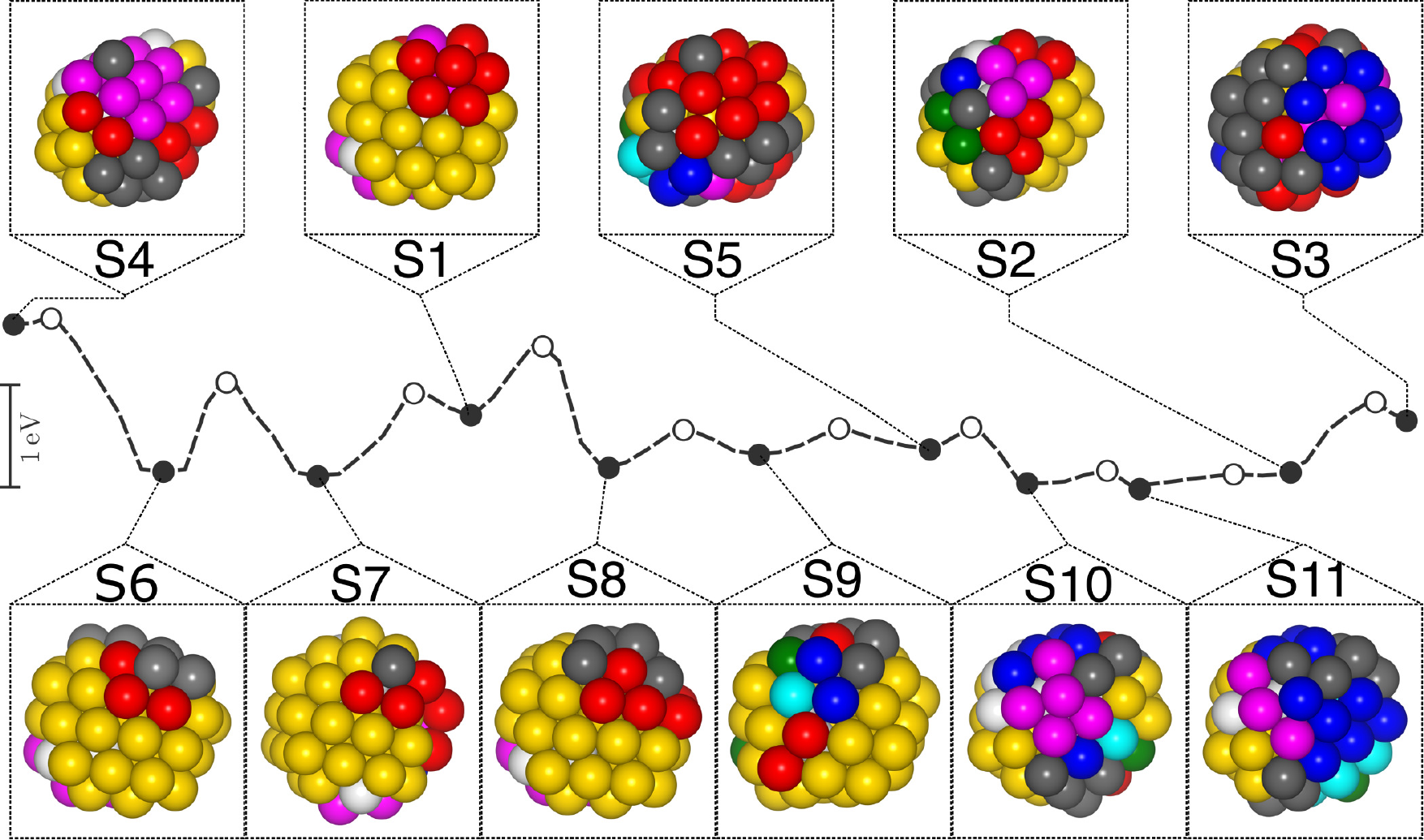}
    \caption{Color coding of the atoms in the Au$_{55}$ structures $S1-S11$ according to the local environment characterized by the three
     CNA pairs, 421, 422 and 555, as in figure 8.
    The minimum energy path connecting the structures is also shown (as in figure \ref{NEB_barriers}). 
    Local energy minima and saddle points are represented by black and white circles, resp.}
    \label{fig:CNA}
\end{figure}

The evolution of the local atomic ordering along the minimum energy path connecting structures $S1-S11$ is shown in figure \ref{fig:CNA}. 
The color scheme is the same as in figure \ref{fig:CNA_example}, 
based on the three types of pairs that describe well the different types of local ordering (\textit{i.e.} 421, 422 and 555). 
Structures $S1$, $S7$, $S8$ and $S9$ in figure \ref{fig:CNA} are predominantly yellow, indicating local icosahedral ordering. 
Again, it can be seen how different two structures can be even if they are separated by only a single low energy barrier. 
A transition involving a hop over a single energy barrier changes the local environment of several atoms. 
The structures obtained for the Au$_{55}$ cluster by fitting the AC-STEM image can be described as `Janus' structures in that 
they have a part that corresponds to icosahedral structure and another part that exhibits local order characteristic of the FCC crystal.
The cluster has two regions, each corresponding to one of these two incompatible structural motifs.\cite{Jonsson_1988}


\subsubsection{Electronic density}


\begin{figure}[!b]
    \centering
    \includegraphics[width=\columnwidth]{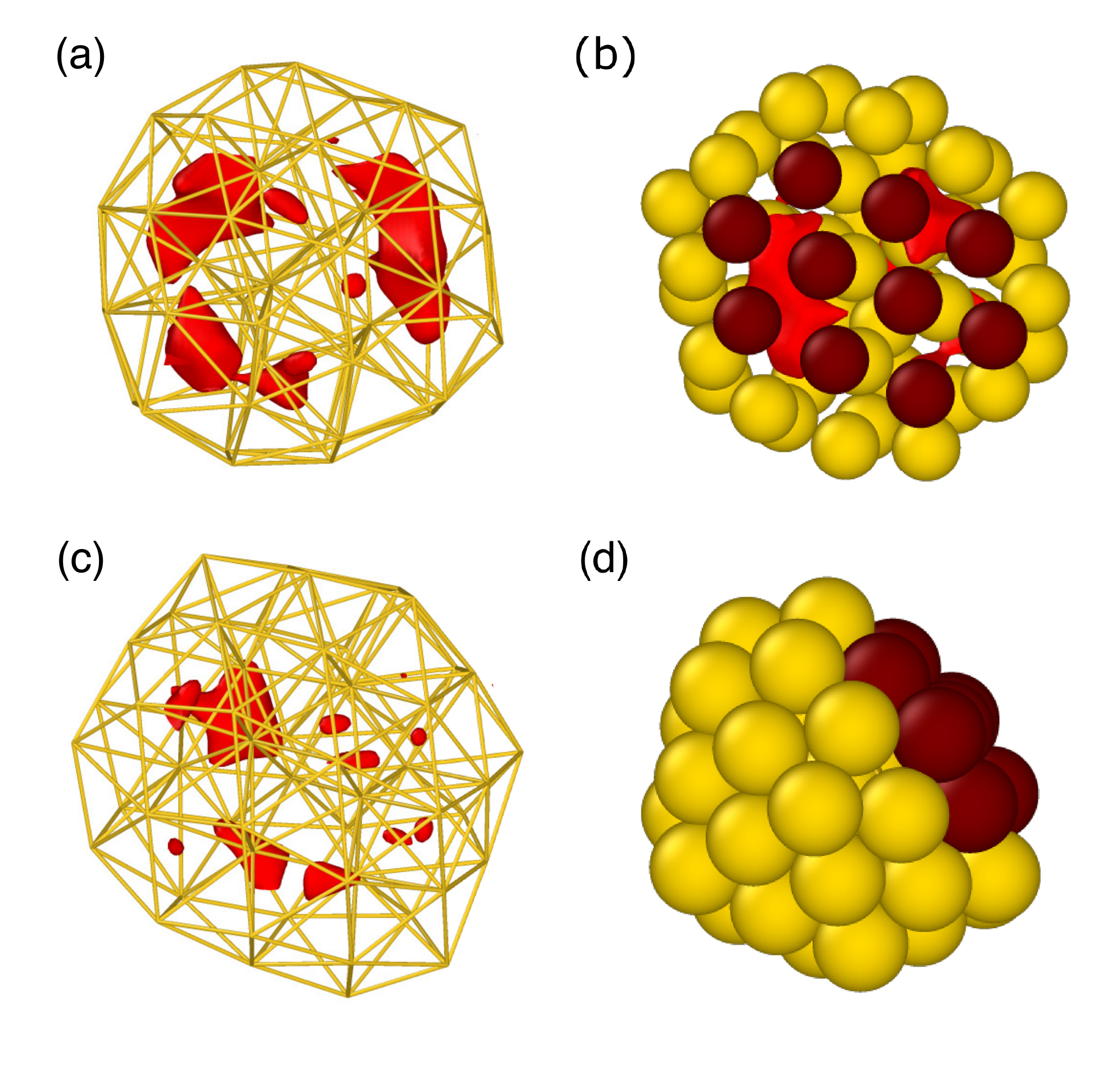}
    \caption{
Electron density isosurface shown in red in (a) and (c) for $S10$ and Garz\'on structures, resp. 
                  The isosurface level rendered corresponds to 0.1 e \AA{$^{-3}$}.
                  This highlights cavities in the electron density between the core of the cluster and the surface shell. 
In (b) and (d) the the atoms forming a flat surface facet in the $S10$ structure are colored in dark red. 
The two figures correspond to different viewing angle.
}
    \label{fig:IsoSurf}
\end{figure}{}

The DFT calculations also give valuable information about the electronic structure of the cluster. 
A surprising feature is that the electron density shows regions of low density 
in between the surface layer of atoms and the inner core of the cluster. 
This is particularly notable for the lowest energy structures, such as $S10$ as shown figure~\ref{fig:IsoSurf}(a). 
The presence of such cavities has been noted previously and discussed in terms of compaction of the surface layer.\cite{VdBossche_2019} 
A more detailed inspection shows that out of the three largest cavities in the structure, two of them are below a flat facet of the cluster. 
Figure \ref{fig:IsoSurf}(b) shows $S10$ with a view directly on this flat facet where the atoms constituting it have been highlighted. 
The atoms are arranged as in a close packed layer, indicating local FCC order. 
While such cavities in the electronic density are also present in the DFT relaxed Garz\'on structure (figure \ref{fig:IsoSurf}(c)),
they are significantly smaller, consistent with a more compact packing of the atoms. 
Figure \ref{fig:IsoSurf}(d) shows an orientation of $S10$ where the flat facet consisting of 10 atoms can be seen clearly. 


\section{Conclusion}\label{sec:Conclusion}

To summarize, a method is presented for performing systematic analysis of experimental AC-STEM images and it is applied 
to the Au$_{55}$ cluster to obtain a set of structures that fit an AC-STEM image quite well. 
A more detailed fit may be impossible in this case as the
structure of such a small and disordered cluster may easily be changing during the AC-STEM measurement. 
The first phase of the method corresponds to the extraction of ($x,y$) coordinates of the atoms by placing a Gaussian representing the contribution of each atom to the image, and assigning a $z$-coordinate using a random number generator to construct a three-dimensional structure. 
Local optimization of the structure is carried out after the placement of each atom using an objective function that combines pixel-by-pixel fit to the image and an approximate energy function to represent the interaction between the atoms. 
Then, after generating several such trial structures using different random number seeds, an optimization is carried out using either GOUST or GA algorithm. An EMT energy function was used in the objective function rather than DFT to keep the computational cost low. 
Even though the lowest energy structure predicted by the EMT potential is inconsistent with the experimental image, it has turned out to be 
adequate as a component in the objective function.
It only serves to give a rough estimate of the distance between neighboring atoms and an attraction to ensure compact structure. 
Finally, the optimized structures are relaxed using electronic structure calculations based on DFT.
A simple numerical measure for the goodness of fit generated as a sum over pixel-by-pixel squared intensity difference
turns out be useful for fine adjustments in the cluster structure, but does not eliminate well enough images that  
differ strongly from the target image. A revised measure where the weight of bright pixels is increased fivefold is found to work better
in such cases.


The potential energy surface in the vicinity of the best fit structures is characterized by calculating the minimum energy path connecting the 
local energy minima.
This reveals the presence of several new, adjacent local minima, some with lower energy and nearly as good fit to the AC-STEM image. 
%
Transitions between neighboring structures along the minimum energy path can in some cases occur readily by thermal activation 
and a hop over a single energy barrier typically involves concerted displacement of several atoms. 
This rearrangement of the atoms can lead to large changes in the simulated AC-STEM images and could shed light on the 
structural fluctuations observed in the AC-STEM experiments. They have been attributed to rotations of the cluster, but from the calculations presented here it is evident that they are more likely due to concerted displacements of several atoms that may even involve a hop over  
a single energy barrier. 
Given that some of the calculated energy barriers between the structures are small, the corresponding transitions could be fast on the time scale of the experimental measurements, so the measured images may be a superposition of two or more structures, making a perfect fit with any
one structure impossible in this case.

The CNA structural analysis performed reveals that the structures obtained can be described as `Janus' type clusters, where a part of the structure has local icosahedral ordering of the atoms and the other part has local order characteristic of the crystal. This is also manifested by a flat
surface segment corresponding to a close-packed layer of atoms, characteristic of the ordering of atoms in the crystal. 

Coordinates of the cluster structures discussed here are listed in the Supporting Information.


\section{Methods}\label{sec:Methods}

\textbf{Minimization of the $\chi^2$ measure.} 
An important aspect of the analysis method presented here is systematic optimization of the atom coordinates as well 
as the width parameter in the Gaussian
representing the signal originating from each atom. Using the $\chi^2$  measure, the steepest descent direction for reducing 
the discrepancy between a simulated and a measured image can be obtained by differentiating $\chi^2$.
Using $I_{ij}\!=\!I_{ij}(\boldsymbol{x},\!\boldsymbol{y},\!\sigma)$ and $\chi^2\!=\!\chi^2 (\boldsymbol{x},\!\boldsymbol{y},\!\sigma)$ 
to shorten the notation, the partial derivative of $\chi^2$ with respect to the $x$-component of atom $k$ can be written as
     \small
       \begin{equation} \label{eq:dx1}
        \frac{\partial {\chi^2}} {\partial{x_k}}= -2 \sum_{i=1}^m\sum_{j=1}^n \left(I'_{ij} - I_{ij}\right) \frac{\partial{I_{ij}}} {\partial{x_k}}, 
    \end{equation}
    \normalsize
    with
     \small
       \begin{equation} \label{eq:dx2}
		\frac{\partial{I_{ij}}} {\partial{x_k}} = \frac {1} {\sigma^2}\left(x'_i - x_k\right) I_{ij}(x_k, y_k,\sigma).
    \end{equation}
    \normalsize
    Substituting eq \ref{eq:dx2} into eq \ref{eq:dx1} and using $ I^{(k)}_{ij}\!=\!I_{ij}(x_k,\!y_k,\!\sigma)$, the $x$-component of the steepest descent vector for atom $k$ is
     \small
       \begin{equation} \label{dx3}
        \frac{\partial {\chi^2}} {\partial{x_k}}= -\frac{2}{\sigma^2} \sum_{i=1}^m\sum_{j=1}^n \left(I'_{ij} - I_{ij}\right) \left(x'_i-x_k\right) I^{(k)}_{ij}.
    \end{equation}
    \normalsize
    Similarly, the $y$-component of the steepest descent vector for atom $k$ is
     \small
       \begin{equation} \label{dy}
        \frac{\partial {\chi^2}} {\partial{y_k}}= -\frac {2} {\sigma^2} \sum_{i=1}^m\sum_{j=1}^n \left(I'_{ij} - I_{ij}\right) \left(y'_j-y_k\right) I^{(k)}_{ij},
    \end{equation}
    \normalsize
and the $\sigma$-component of the gradient vector is 
    \small
       \begin{equation} \label{dsigma}
        \frac{\partial{\chi^2}} {\partial{\sigma}}\!=\!-\frac{2}{\sigma^3}\!\sum_{i=1}^m\sum_{j=1}^n\!\left(I'_{ij}\!-\!I_{ij}\right)\!\left(\left(x'_i\!-\!x_k\right)^2\!+\!\left(y'_j\!-\!y_k\right)^2\right)I_{ij}. \end{equation}
    \normalsize
It is important to note that $I_{ij}\!=\!I_{ij}(\boldsymbol{x}, \boldsymbol{y}, \sigma)$ in the above equations represents the intensity from all atoms whereas $I_{ij}^{(k)}\!=\!I_{ij} (x_k, y_k, \sigma)$ represents the intensity from atom $k$ only. 
    

\textbf{GOUST Algorithm.} 
The method is based on finding new local minima of the objective function by identifying first order saddle points (SPs) on the objective function surface starting from a given local minimum and then sliding down to adjacent local minima on the other side of the SP. 
The method is implemented in the EON software\cite{Chill_2014} and a more detailed explanation of the method can be found in refs. \cite{Pedersen_2010,Plasencia_2014}
To minimize the objective function given in eq \ref{eq:objective-function}, the first step is to perform local minimization starting from some value of the parameters (\textit{i.e.} atom coordinates and Gaussian width) so as to make sure that the search for saddle points will start from a local minimum. 
In this and other energy minimization calculations carried out here, the velocity projection optimization (VPO) method\cite{Jonsson_1998} is first 
used until the magnitude of atomic forces has dropped below 0.5 eV/{\AA}, and then convergence to 0.01 eV/{\AA} is obtained with the BFGS \cite{John1985,NoceWrig06} method. 
After a local minimum has been reached, a slight change in the parameter values (here atom coordinates and Gaussian width) is made,
as in the adaptive kinetic Monte Carlo method.\cite{Pedersen_2011}
The set of new values serves as a starting point for the SP searches. A climb up the objective function surface to a SP is carried out using the minimum mode following method.\cite{Henkelman_1999,Gutierrez_2017} 
After reaching a SP, a displacement along the minimum mode vector is made followed by local minimization to get to the adjacent minimum. 
For a given local minimum, several SP searches are carried out in order to find several surrounding SPs and adjacent minima. Finally, from the list of local minima found in this way, the ones with the low value of the objective function are selected and a new searches for SPs carried out 
from there. 
The process is repeated until the system reaches a local minimum of the objective function that is lower than any of the nearby local minima.
In this way, the system is taken down a funnel on the objective function surface. 

\textbf{Genetic Algorithm.} 
A genetic algorithm (GA) implemented in the Atomic Simulation Environment (ASE) \cite{ase-paper:2017, BahnJacobsen:2002} is used. The starting configurations were generated by the fitting process in the first phase. During the GA evolution, new structures are generated either by crossover or mutation. In the case of crossover, two structures are selected from the population and an offspring is produced through the pairing of the parent structures. Crossover is carried out $75\%$ of the time using the cut-splice algorithm \cite{Wolf:1998}. To maintain population diversity, mutation is performed $25\%$ of the time. During a mutation a cluster is selected and a new structure is generated by randomly displacing the atoms 
by a distance that ranges from 0 to $\SI{1.5}{\angstrom}$. The new structure undergoes local minimization of the combined objective function and the population is updated so that it always contains the best 20 unique structures. The loop is repeated until 2000 structures have been tested.
	
\textbf{DFT Calculations.} 
The DFT calculations are performed using the PBEsol functional approximation\cite{Perdew_2008} 
and a plane-wave basis set with a kinetic energy cutoff of $\SI{250}{\electronvolt}$. 
The choice of this functional is motivated by recent studies where PBEsol is found to be more accurate than other generalized gradient approximation functionals for small gold clusters \cite{Johansson:2008}. 
The calculations are performed in a cubic simulation box of length \SI{25}{\angstrom} on each side.
Gaussian smearing is used with a width of \SI{0.1}{\electronvolt}. Spin polarization is not included. 
The Vienna {\it ab initio} simulation package (VASP) is used in these calculations.\cite{Kresse:1996} 
    
\textbf{CI-NEB Calculations.} 
The minimum energy path between structures $S1$-$S5$ is calculated using the CI-NEB method.\cite{Henkelman_2000a,Henkelman_2000b,Asgeirsson_2020} 
Six intermediate images are used to provide a discrete representation of the path. The image dependent pair potential (IDPP) method\cite{Smidstrup_idpp_2014} is used to generated an initial path from which the CI-NEB optimization is carried out. 
To avoid rotation of intermediate images, the RT-NEB algorithm is used.\cite{Melander_2015}


\textbf{Preparation of AC-STEM Image.} 
The experimental AC-STEM image used in this work is adapted from figure 3(a) in ref.\cite{Wang2012b}. 
Based on the scale bar presented in the image, 
a box of size roughly $\SI{2.5}{\nm}$ by $\SI{2.5}{\nm}$ containing the cluster is cropped from the figure.
Although the image had been processed by the authors to remove noise and enhance image quality, some noise is still visible 
in the vicinity of the cluster, especially around the edges. 
For example, small bright spots spatially separated from the main cluster can be seen in the image. 
These are indicated by red circles in figure S3 in the Supporting Information.  
These isolated features and small intensity patches around the edges are not considered as part of the cluster and are 
removed from the image. 
Finally, Gaussian averaging of the pixel intensity is used to generate the AC-STEM image shown in figures 1-3.
%

    
\section{Acknowledgements}
This project was funded by the Icelandic Research Fund and the research fund of the University of Iceland. 
KSB acknowledges a Ph.D. fellowship from the Doctoral Fund of the University of Iceland. 
We thank Emil Gauti Fridriksson for help in implementing the objective function in the EON software.  
The calculations were carried out at the Icelandic Research High Performance Computing (IRHPC) facility.


\bibliography{ref}

\end{document}